\documentstyle[psfig]{mn2e}

\def\gs{\mathrel{\hbox{\rlap{\hbox{\lower4pt\hbox{$\sim$}}}\hbox{$>$}}}}
\def\ls{\mathrel{\hbox{\rlap{\hbox{\lower4pt\hbox{$\sim$}}}\hbox{$<$}}}}

\title[Long-term spectral changes in 1H~0707--495]
{
Long-term spectral changes in the partial-covering candidate NLS1 1H~0707--495
 }

\author[L. C. Gallo et al.]
{L. C. Gallo,$^1$ Y. Tanaka,$^1$ Th. Boller,$^1$ A. C. Fabian,$^2$ S. Vaughan,$^3$ and W. N. Brandt$^4$  \\
$^1$ Max-Planck-Institut f\"ur extraterrestrische Physik, Postfach 1312, 85741 Garching, Germany \\
$^2$ Institute of  Astronomy, Madingley Road, Cambridge CB3 0HA\\
$^3$ X-ray and Observational Astronomy Group, Department of Physics and Astronomy, University of Leicester, Leicester LE1 7RH \\
$^4$ Department of Astronomy and Astrophysics, The Pennsylvania State University, 525 Davey Lab, University Park, PA 16802, USA\\
}

\date{Accepted. Received. }

\begin{document}
\label{firstpage}
\maketitle

\begin{abstract}
We compare two {\em XMM-Newton} observations
of the Narrow-Line Seyfert~1 galaxy 1H~0707--495 separated by two
years, and discuss the results in terms of the partial-covering phenomenon.
The second longer observation once again displays a sharp ($<~200$ eV) 
spectral drop above 7~keV; however, in comparison to the first observation, 
the edge depth and energy have changed significantly.
In addition to changes in the edge parameters, the high-energy spectrum 
appears steeper.  
The changes in the high-energy continuum can be adequately explained
in terms of a partial-covering absorber outflowing from the
central region.
The low-energy spectrum also shows significant long-term spectral variability,
including: a substantial increase in the disc temperature; detection of
a $\sim 0.9$~keV emission feature; and the presence of warm absorption
that was also detected during the $ASCA$ mission, but not seen during the first
{\em XMM-Newton} observation.  
The large increase in disc temperature, and more modest rise in luminosity,
can be understood if we 
consider the slim-disc model for 1H~0707--495. 
In addition, the higher disc luminosity could be the 
driving force behind the outflow scenario and the re-appearance of a 
warm medium during the second {\em XMM-Newton} observation.

\end{abstract}


\section{Introduction}
The first observation of the Narrow-Line Seyfert~1 galaxy (NLS1) 1H~0707--495 ($z = 0.0411$) with {\em XMM-Newton} in October 2000 revealed
a sharp spectral feature at $\sim 7$~keV in which the spectrum ``jumped'' by a factor
of $> 2$ within a few hundred eV (Boller et al. 2002; hereafter B02).
The measured energy and sharpness of the feature implied K-absorption by
neutral iron; however, the absence of a fluorescence line suggested a more
complicated situation.  These observations coupled with the low intrinsic
absorption below $\sim 1$~keV led B02 to consider a partial-covering model 
(Holt et al. 1980) to interpret the X-ray spectrum.  Although a good fit was obtained, 
the model required a steep intrinsic power-law ($\Gamma = 3.5$) and an iron overabundance
of $> 35$ times solar.

In an alternative approach, Fabian et al. (2002) treated the spectrum as 
being dominated by a 
reflection component, where the deep edge was interpreted as the blue wing 
of a relativistically broadened iron line.  This model was also successful in describing
the 0.5--11~keV spectrum (including the soft-emission component), and required a 
more modest iron overabundance of 5--7 times solar.

A second attempt at fitting the spectrum with a partial-covering model was made by
Tanaka et al. (2004; hereafter T04).  They 
demonstrated that a partial-covering model, implementing a cut-off power-law (rather 
than a simple power-law), greatly reduced the iron abundance 
to $\approx 5$ times solar. 

Three months following the {\em XMM-Newton} observation, 
1H0707--495 was observed 
with $Chandra$ (Leighly et al. 2002; hereafter L02). The
flux was found to be about 10 times higher than what was measured with
{\em XMM-Newton}. 
A preliminary investigation found that
the edge feature was weak, possibly even absent,
during the $Chandra$ observation.

1H~0707--495 was observed with {\em XMM-Newton} for a second time two
years after the first observation.  
In this paper we will focus on the differences between the two 
{\em XMM-Newton} observations,
and attempt to explain them in the context of the partial-covering phenomenon.
A reflection interpretation is also possible, and  
is investigated further by Fabian et al. (2004).

\section{Observations and data reduction}
The second {\em XMM-Newton} observation of 1H~0707--495 occurred on 2002 
October 13 during revolution 0521 (hereafter this observation will be referred to as AO2).
The total duration was 80 ks, during which time all instruments were 
functioning normally.  The observation was carried out identically to the
first Guaranteed Time observation during revolution 0118 (hereafter
referred to as GT).  
The EPIC cameras were operated in full frame mode and utilised the medium
filter.

The Observation Data Files (ODFs) from the GT and AO2 observations
were processed to produce calibrated event lists using the {\em XMM-Newton} 
Science Analysis System ({\tt SAS v5.4.1}).
Unwanted hot, dead, or flickering pixels were removed as were events due to 
electronic noise.  Event energies were corrected for charge-transfer 
inefficiencies.  EPIC response matrices were generated using the {\tt SAS} 
tasks {\tt ARFGEN} and {\tt RMFGEN}.  Light curves were extracted from these 
event lists to search for periods of high background flaring.  
A background 
flare was detected during the first few ks of the AO2 observation, and the data
have been ignored during this interval.  
The total amount of good 
exposure time selected was 70 ks and 78 ks for the pn and MOS detectors, 
respectively.
Source photons were extracted from a circular region 35$^{\prime\prime}$ across
and centred on the source.
The background photons were extracted from an off-source region and 
appropriately scaled to the source selection region.
Single and double events were selected for the pn detector, and 
single-quadruple events were selected for the MOS.  
The total number of source counts collected in the 0.3--10~keV range by
the pn, MOS1, and MOS2 were 275451, 46284, and 46426, respectively.
Comparing the source and 
background spectra we found that the spectra are source dominated below 10~keV.

High-resolution spectra were obtained with the Reflection Grating Spectrometers (RGS).  The RGS were operated
in standard Spectro+Q mode, and collected data for a total of 76 ks.  The
first-order RGS spectra were extracted using the {\tt SAS} task {\tt RGSPROC}, and the response matrices were generated using {\tt RGSRMFGEN}.  The RGS data were background dominated during the GT observation.

\section{The AO2 spectral and timing analysis}

\subsection{Spectral Analysis}

The
source spectra were grouped such that each bin contained at least 40    
counts.
Spectral fitting was performed using {\tt XSPEC v11.2.0} (Arnaud 1996).  Fit
parameters are reported in the rest-frame of the object.
The quoted errors on the model parameters correspond to a 90\% confidence
level for one interesting parameter (i.e. a $\Delta\chi^2$ = 2.7 criterion).
The Galactic column density toward 1H~0707--495 is $N_H$ = 5.8 $\times$
10$^{20}$ cm$^{-2}$ (Dickey \& Lockman 1990).
Element abundances from Anders \& Grevesse (1989) are used throughout.
Luminosities are derived assuming isotropic emission and a standard
cosmology with $H_0$=$\rm 70\ km\ s^{-1}\ Mpc^{-1}$, 
$\Omega_{M} = 0.3$, and $\Omega_\Lambda = 0.7$.

While both the pn and MOS data from each observation were examined for
consistency, the discussion focuses on the pn results due to the higher
signal-to-noise and the better stability of the pn calibration over the two
years separating the observations.

\begin{figure}
       \psfig{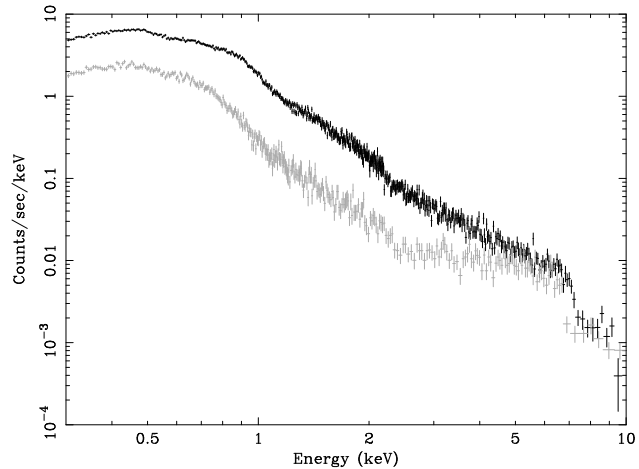}
      \caption{The 0.3--10~keV AO2 (black data points) and GT
(gray data points) pn data separated by about 2 years.
}
\label{data}
\end{figure}

Figure~\ref{data} gives a direct comparison between the GT and AO2 
time-averaged spectra. 
In general, the count rate below $\sim$ 5~keV is higher during the AO2
observation; however, between 5--10~keV the fluxes are comparable.
The flattening of the spectrum above $\sim$ 2~keV, seen in the GT data,
is clearly diminished in the AO2 data.  The depth ($\tau$) and energy ($E$) of
the edge-like feature at $\sim$ 7~keV have changed.

Fitting a power-law and edge to the 3--10~keV AO2 data resulted in a good fit
($\chi^2$ = 126 for 122 $dof$), with the edge parameters $E = 7.49 \pm 0.10$~keV
and $\tau = 0.84^{+0.25}_{-0.22}$.  In addition the edge was sharp, with an
intrinsic width of $< 160$~eV (measured with the {\tt SMEDGE} (Ebisawa 1991)
model in {\tt XSPEC}).  
Treating the $\sim 7.5$~keV edge
as a blueshifted neutral edge, we would expect neutral iron emission at
$\sim 6.7$~keV.  However, the best-fit line ($E \approx  6.74$~keV)
was broad ($\sigma \approx 229$ eV) and only a marginal improvement to the 
edge fit ($\Delta\chi^2 = 6$ for 3 additional free parameters).
The broadness of the line is problematic, as it is expected that the
line would be narrower than the edge.

The 90\% upper limit on the line flux is $1.08 \times 10^{-14}$ erg s$^{-1}$ 
cm$^{-2}$, whereas the flux absorbed by the edge is 
$8.49 \times 10^{-14}$ erg s$^{-1}$, resulting in a line-to-edge flux
ratio of 0.13.  For a spherically symmetric distribution of absorbing material, 
the expected line-to-edge flux ratio is approximately equal to the 
fluorescent yield for iron (0.34; Bambynek et al. 1972).  Therefore, for
1H~0707--495 the expected
ratio is more than 2.5 times the measured value, and the discrepancy
is even larger if the edge and line are treated as arising from ionised iron
(Krolik \& Kallman 1987).
The measured line-to-edge flux ratio indicates that the absorber covers a
solid angle $\Omega/4\pi \ls 0.4$.

When applying the partial-covering model used by T04 to interpret the GT 
observation 
to the AO2 data it was found that the two data sets were very similar.
The 2--10~keV spectrum could be fitted with a cut-off power-law 
($\Gamma \approx 2$, $E_c \approx 5$~keV), modified by an absorber with an iron overabundance of 
$\sim 5$ times solar, 
consistent with what was used to model the GT data.  
The primary difference
is that the AO2 model only requires one absorber, which is equivalent to 
setting 
the covering fraction of the second absorber, used on the GT data, to zero.
The reduced amount of absorption can simultaneously explain the apparently 
steeper spectrum and the shallower edge seen in the AO2 data.
Indeed, the intrinsic, unabsorbed 2--10~keV spectra from the two observations 
were entirely consistent, with a flux difference of only a factor
of $\sim 2$.
The edge was found at a higher energy (7.5 $\pm$ 0.1~keV) during the AO2
observation, and this is discussed at length in Section~4.
The use of the power-law with a low cut-off energy is empirical.
Allowing for curvature in the high-energy spectrum greatly reduces the 
iron abundances.

Considering the broad-band (0.3--10~keV) spectrum, a multi-coloured disc
black body component (MCD; Mitsuda et al. 1984; Makishima et al. 1986) was 
included to fit the soft-excess below $\sim 1.5$~keV.  

In treating the intrinsic absorption in the GT data, T04 found
that including an edge at 0.37~keV (0.39~keV rest-frame) was an improvement
over using only neutral absorption.  It was uncertain if the edge was required
due to calibration uncertainties or if it was due to intrinsic C~\textsc{v}
absorption in 1H~0707--495.  In order to make the most direct comparison with
the earlier observation we also included an edge at 0.37~keV, in addition
to Galactic and intrinsic cold absorption.  All model
parameters for the complete AO2 fit are given in Table~1. 

While the continuum model of an intrinsically absorbed MCD and cut-off 
power-law modified by partial-covering provided a good fit to the high- and
low-energy spectra, the fit was inadequate in the 0.5--2~keV range 
($\chi^2$ = 1340 for 569 $dof$).
The AO2 continuum appeared to be modified by line-like
absorption and emission.  Fits using only one
Gaussian profile (either an absorption or emission line) were unsuccessful,
as were attempts to fit the data with an absorption edge.
Two distinct Gaussian profiles, one absorption and one emission, were required.
An absorption profile
with an equivalent width of $EW \approx -50$ eV, an energy of
$E \approx$ 1.17 keV, and a width of $\sigma \approx 115$~eV,
as well as an emission profile with $EW \approx$ 25 eV, $E \approx$ 0.92~keV,
and $\sigma \approx 50$~eV,
were included in the broad-band fit ($\Delta\chi^2$ = 684 for 6 addition 
free parameters).

The RGS spectra are of low signal-to-noise, but confirmed the existence of 
the absorption feature detected in the pn.
The improvement to the partial covering continuum, by the addition of the
Gaussian absorption profile was $\Delta\chi^2 = 231$ for 3 additional free
parameters.  The line energy and width was
$E = 1.19^{+0.02}_{-0.11}$~keV and $\sigma = 245^{+133}_{-32}$~eV, respectively.  
No strong emission features were prominent in the RGS
spectra at approximately 0.92~keV (only one RGS is effective in this energy
range); however there was indication of a skewed excess at a slightly lower
energy which could be consistent with the EPIC findings. 
The absence of a convincing emission feature in the RGS is probably a result 
of the poorer signal-to-noise, since the feature was detected in all three 
EPIC instruments, as well as the earlier $Chandra$ observation (L02).

\begin{figure}
       \psfig{figure=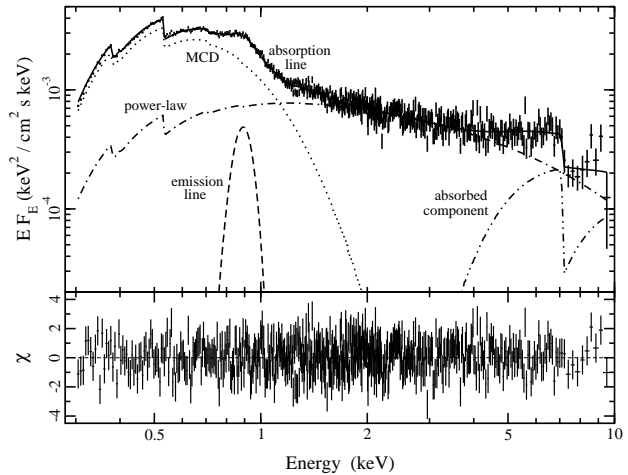,angle=-90,width=9.0cm,clip=}
      \caption{The best-fit partial-covering model fitted to the AO2 EPIC
pn data.
In the upper panel we present the unfolded model with
the individual components marked. The fit residuals (in terms of sigma)
are shown in the lower panel. 
}
\label{fit}
\end{figure}

The complete partial-covering model applied to the EPIC pn AO2 data is
presented in Figure~\ref{fit} ($\chi^2$ = 656 for 563 $dof$),
and the fit parameters are given in Table 1.  The 0.3--10~keV flux and
luminosity, corrected for Galactic absorption, are 
$1.1 \times 10^{-11}$ erg s$^{-1}$ cm$^{-2}$ and $4.4 \times 10^{43}$ erg 
s$^{-1}$, respectively.  The 2--10~keV luminosity is $4.7 \times 10^{42}$
erg s$^{-1}$.
\begin{table*}
\caption[]{
Model parameters for the partial-covering fit to the pn data
($\chi^2 = 656$ for 563 $dof$).  
Superscript $f$ indicates that the parameter 
is fixed.  Fluxes ($F$) are given in units of erg s$^{-1}$ cm$^{-2}$ and
have been corrected for Galactic absorption.
}
\label{tab1}
\begin{center}
\begin{tabular*}{6.0in}{l c c c c }\hline\hline
Low-Energy Absorption &  &  &  &        \\
 & Galactic$^{f}$ & 5.8 $\times$ 10$^{20}$ cm$^{-2}$ &  &       \\
 & Intrinsic (cold) & (1.6 $\pm$ 0.2) $\times$ 10$^{20}$ cm$^{-2}$ &  &  \\
 & Intrinsic (edge) & $E^{f} = 0.37$ keV & $\tau = 0.37 \pm 0.05$ &  \\
Continuum & & &  & \\
 & MCD & $kT = 155 \pm 1$ eV & & $F = 1.00 \times 10^{-11}$  \\
 & Cutoff power-law & $\Gamma = 2.0 \pm 0.1$ & $E_{c} = 4.5^{+0.9}_{-0.7}$ keV & $F = 4.11 \times 10^{-12}$ \\
Partial Coverer &  &  &  & \\   
 & Absorption & $N_{H} = 43^{+14}_{-19} \times 10^{22}$ cm$^{-2}$ &  & \\
 & Fe abundance & $3^{+7}_{-1} \times$ solar &  & \\
 & Covering fraction & 0.69 $\pm$ 0.01 & &  \\
Line Features &  &  &  & \\   
 & Emission & $E = 921 \pm 10$ eV & $\sigma = 50 \pm 13$ eV & $EW \approx 25$ eV \\
 & Absorption & $E = 1.17^{+0.02}_{-0.04}$ keV & $\sigma = 114^{+29}_{-17}$ eV & $EW \approx -50$ eV \\
\hline\hline
\end{tabular*}
\end{center}
\end{table*}

\subsection{Timing Analysis}
\begin{figure}
       \psfig{figure=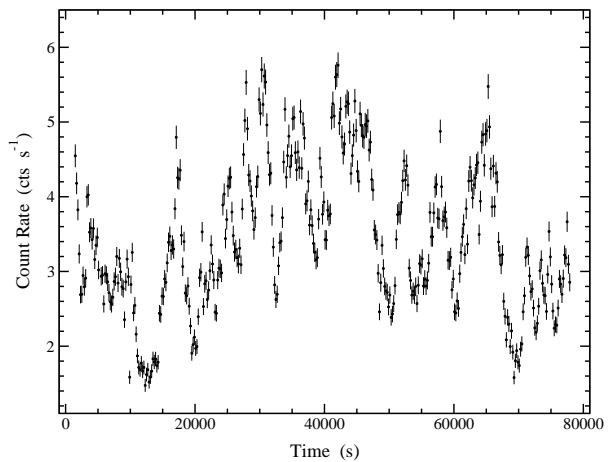,angle=-90,width=8.3cm,clip=}
      \caption{The AO2 pn light curve in the 0.3--12 keV band.  Bin sizes are
200~s.
}
\label{lc}
\end{figure}

The 0.3--12~keV pn light curve from AO2 is presented in Figure~\ref{lc}.  
The average count rate was about three times higher than it was during the
GT observation.  Count rate variations by about a factor of four were
seen throughout the observation.  The persistent
and rapid variability is quite typical of 1H~0707--495 (Leighly 1999; B02).

\begin{figure}
       \psfig{figure=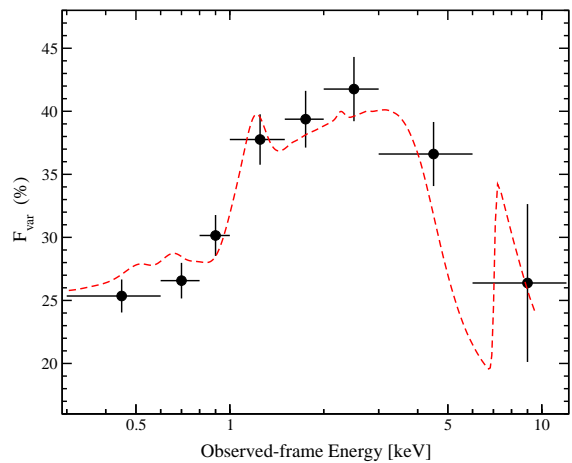,angle=-90,width=8.3cm,clip=}
      \caption{
The F$_{var}$ calculated in eight energy bins between 0.3--12~keV
using 400~s binning of the light curves (data points).
Overplotted is the
ratio of the high- and low-flux spectral models (red,
dashed line).
}
\label{fvar}
\end{figure}

During the GT observation, 1H~0707--495 displayed no significant spectral
variability.  Hardness ratio curves showed some variability (Figure 5 of B02),
but they were uncorrelated with flux.  The fractional variability amplitude
F$_{var}$, (Edelson et al. 2002; using the unweighted mean count rate and 
sigma), was calculated in six energy bands between
0.1--10 keV to examine the degree of variability in each band; it was
determined to be constant during the GT observation (Figure 4 of B02).  
However, the F$_{var}$ spectrum portrays substantially more spectral 
variability during the AO2 observation (data points in Figure~\ref{fvar}).

Figure~\ref{fvar} appears similar to the rms spectrum
from the high-flux state $Chandra$ observation (L02), as well as to the rms
spectrum of some broad-line Seyfert~1s, such as NGC~4151 (Zdziarski et al. 
2002).
In addition, Figure~\ref{fvar} is
notably similar to the F$_{var}$ spectrum
of another
partial covering candidate NLS1, IRAS~13224--3809 (Gallo et al. 2004). 

Fabian et al. (2004) modelled the rms spectrum shown in Figure~\ref{fvar}
and demonstrated that it could be well-fitted as the superposition of two 
components with different intrinsic rms:  a variable
power-law (in flux only), and a less variable reflection component.

The energy-dependent variability can also be explained within the
partial covering scenario.  For a finite extension of the emission
region, the covering condition can be more complex than described by a
single absorber with parameters $f$ (covering fraction) and $N_{H}$
(column density in units of $10^{22}$ cm$^{-2}$).  
As a better approximation we assume two partial
coverers (double partial covering; see T04), and then model the
high-flux ($> 5$~cts s$^{-1}$) and low-flux ($< 3$~cts s$^{-1}$) spectra
simultaneously for a common intrinsic spectrum (e.g. see Figure~7 of 
Boller et al. 2003).  
The high- and low-flux spectra are grouped with bins containing at least 40 and
20 counts, respectively.
All model parameters of the high- and low-flux spectra are linked, and only
the covering fraction ($f$) of each absorber is allowed to vary 
independently.
The model gives a satisfactory fit ($\chi^2 = 770.9$ for 683 $dof$).
The covering parameters of the two absorbers ($f,~N_{H}$) are determined to 
be (0.61, 36.0) and (0.46, 0.21) in the high-flux state, and
(0.80, 36.0) and (0.0, 0.21) in the low-flux state.
In Figure~\ref{fvar} we show the ratio (high/low) of the two models (dashed-line)
overplotted on the rms spectrum.  A double partial covering can
reproduce the energy-dependent variations in shape and amplitude.

\section{Discussion}

\subsection{A shifting edge and relativistic outflows}
\begin{figure}
       \psfig{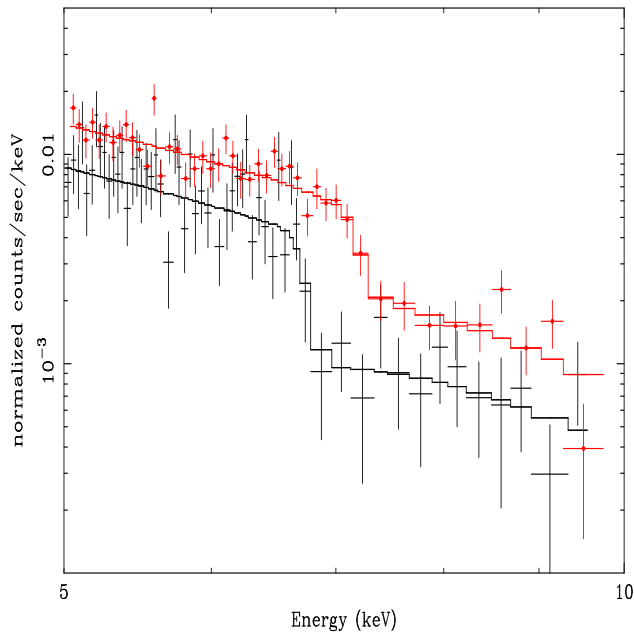}
      \caption{The edge plus power-law fitted to the AO2 (top curve; red dots)
and GT (bottom curve; black crosses) pn data.
The edge energy and depth are clearly different; however, the edge width
remains comparable (and narrow).
}
\label{edge}
\end{figure}

Perhaps the most interesting changes between the two observations are
the differences in the edge parameters (Figure~\ref{edge}). 
During the GT observation the edge energy was $E = 7.0 \pm 0.1$~keV, 
entirely consistent with neutral iron.
At the second epoch the edge energy was considerably higher,
$E = 7.5 \pm 0.1$~keV.  In addition, the edge depth during AO2 was 
approximately half as deep as it was during the first observation.
The intrinsic edge widths were narrow at both epochs.  The 90\% confidence
upper-limits on the intrinsic edge widths were 350~eV and 160~eV during
the GT and AO2 observations, respectively.
If we continue to adopt the iron edge interpretation for the second
observation, the edge energy 
corresponds to absorption by Fe~\textsc{vii}--\textsc{x}.

Two scenarios emerge as possible explanations for the differing edge
energies: (1) a change in
the ionisation state, or (2) a high-speed outflow.

While a change in the
ionisation state due to a change in the ionising continuum is plausible,
the drawback is that the width of the edge is narrow in both observations.
Palmeri et al. (2002) discuss two distinct spectral features of the
Fe K edge which seem relevant here.  First, a sharp edge is only expected in 
low-ionisation plasmas ($\xi$ = ($L/nR^2$) $\ls 10$ [cgs units are hereafter 
assumed]; where $L$ is the
incident X-ray luminosity, $n$ is the gas density, and $R$ is the distance from
the radiation source), in which case the edge energy would be consistent with
neutral iron within the available resolution.  Secondly, for higher ionisation 
plasmas ($\xi$ $\ge$ 1), a strong absorption feature appears at 
about 7.2~keV, arising from the K$\beta$ unresolved transition array.
Even within our uncertainties, we can dismiss absorption features below
$\sim$ 7.3~keV; therefore a photoionisation interpretation for the shifting
edge seems unlikely.

On the other hand,
if the edge arises from a neutral absorber, the observed energy implies
an approaching velocity of ($0.05 \pm 0.01$)$c$.  This interpretation is consistent
with the edge width remaining constant over the two year span.
In the context of the partial-covering phenomenon, assuming that
new absorbers are
not being generated, the outflowing material would result in a decrease of
the covering fraction, and subsequently a decrease in the edge depth as the 
absorbers cover a smaller solid angle.
Both effects, a shallower edge and reduced absorption, are observed in the 
AO2 data.  Evidence for an outflow was also seen in the $HST$ spectrum of
1H~0707--495 (Leighly 2004), although not at such high velocities. 
An equally sharp feature in IRAS~13224--3809 (Boller et al. 2003)
was found at $8.2 \pm 0.1$~keV.  The outflow scenario was also suggested 
for that NLS1. 

Considering a patchy-absorber model, and following, for example,
Turner et al. (1993), Reynolds (1997) and Ogle et al. (2004), 
the mass loss rate can be 
estimated by combining the continuity equation  
$\dot{M} = 4\pi R^2 f_v n m_p v$ (where $f_v$ is the volume-filling 
factor and $n$ the atomic density) with the expression for the ionization 
parameter to obtain $\dot{M} = 4\pi f_v (L/\xi) m_pv$. 
For $v=0.05c$, ionizing luminosity $L=5\times10^{43}$ erg s$^{-1}$ and $\xi \ls 1$, 
this gives $\dot{M} \gs 10^4 (f_v/\xi)$ M$_{\odot}$ yr$^{-1}$ 
and a kinetic luminosity $L_w \gs 10^{48} (f_v/\xi)$ erg s$^{-1}$. 
The mass accretion rate estimated in TO4, assuming a slim disk (see Section 
4.2.3) 
and their estimated mass, is $\dot{M}_{acc}\approx 0.1$ M$_{\odot}$ yr$^{-1}$. 
For the outflow rate to be consistent with the mass accretion rate, 
$f_v/\xi < 5\times 10^{-6}$ [cgs] is required, i.e. the absorber is extremely clumped. 
A clumped medium is also the simplest way to reduce the mass flux
(Krolik et al. 1985).

Inserting the value of $L$ into $\xi = L/nR^2$ gives $nR^2 = 5\times10^{43}/\xi$ 
[cgs], where $R$ is the distance to the absorber. For simplicity, we assume  
spherical ``blobs" of radius $r$, which are randomly distributed. The covering 
fraction obtained from the fit shows that the average number of intersecting 
blobs in the line-of-sight is $\sim2$ (Tanaka, Ueda, Boller 2003). Using the 
column density estimated from the fit, $4nr=4\times10^{23}$ cm$^{-2}$, we get 
$R^2/r=5\times10^{20}/\xi$ cm. 
The mean free path of an ionizing photon is given by $l= 4r/3 f_v$. 
Since $R > 2l$; 
$R^2/r > 8R/3f_v$,  
we obtain $R<2\times10^{20}(f_v/\xi) \sim 10^{15}$ cm, 
and $r < 2\times10^{9}\xi$ cm. These values are not inconsistent with those 
estimated from the time scale of variability (TO4).
The atomic density is very high, $n > 5\times10^{13}/\xi$ cm$^{-3}$.
Such blobs would need to be confined by magnetic fields (e.g. Rees 1987).

\subsection{The low-energy spectrum 
}
At both epochs, the low-energy continuum is dominated by a thermal
component which can be successfully fitted with a MCD.  
The inner-disc temperature is noticeably higher ($\sim 50$\%) during AO2
($kT_{AO2} \approx$ 155~eV, $kT_{GT} \approx$ 103~eV).

\begin{figure}
       \psfig{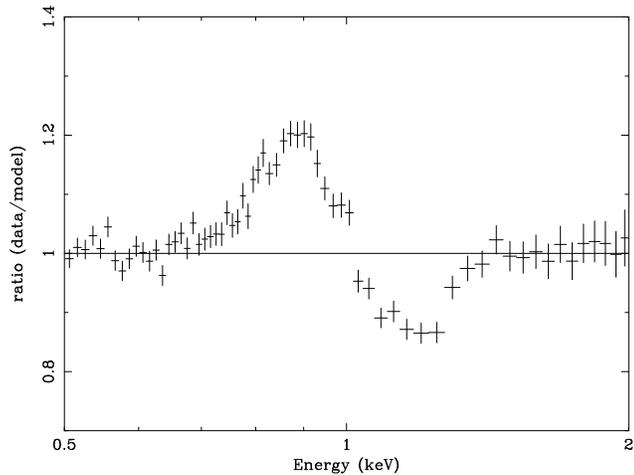}
      \caption{Spectral complexity in the low-energy spectrum.  The residuals
(data/model) remaining between 0.5--2~keV after the AO2 pn spectrum is fitted
with a MCD plus power-law model (see text for details). 
}
\label{moratio}
\end{figure}

In addition, a strong ``P Cygni-type'' type feature is also observed during 
the AO2 observation (Figure~\ref{moratio}).
The profile is likely coincidental and unrelated to the outflow scenario
since there is no evidence of P Cygni-type profiles in the RGS.  

\subsubsection{An extended warm medium
}
Qualitatively, the P Cygni-type feature is similar to that detected during 
the $Chandra$ observation of 1H~0707--495.
In a preliminary investigation L02 determined that
the feature was best described when both emission and absorption were
considered.
We have also treated the feature as a blended absorption and emission profile
from a warm medium.

Similar absorption features have been observed in other NLS1.
Nicastro, Fiore \& Matt (1999) suggest that such absorption features are 
likely a
blend of resonance absorption lines due to mostly ionised L-shell iron. 
Interestingly, IRAS~13224--3809, another partial-covering NLS1
candidate (Boller et al. 2003), also displays a similar absorption feature.

The emission feature can be explained as arising from an extended warm
medium, at lower ionisation, and lying outside the line-of-sight.
In addition to the $Chandra$ observation, similar emission features have been 
observed 
in the NLS1:
NGC~4051 during its
low-flux state (Uttley et al. 2003), and Mrk~1239 (Grupe et al. 2004).  
For both objects, the authors attributed the emission to ionised Fe and/or
Ne.
In a thorough investigation of the low-state RGS spectra of NGC~4051
Pounds et al. (2004) identified emission features around 900~eV as arising
from Ne~\textsc{ix} and the radiative recombination continua from O~\textsc{viii}.
Detailed simulations of high-resolution data are still required to 
address the exact nature of this feature in 1H~0707--495.

The emission/absorption feature was not detected during the low-flux GT observation (B02; Fabian et al.
2002; T04).  Simulations of the pn data using the AO2 model and a 40~ks 
exposure time (consistent with the GT exposure time) indicated that both
the emission and absorption features should have been significantly detected 
during the shorter GT observation, if they were present.
Furthermore, the absorption feature was detected in the $ASCA$
observations of 1H~0707--495 (Leighly et al. 1997; Leighly 1999; Vaughan et
al. 1999)
when the flux was comparable to the AO2 flux. 
Indications are that these warm features are time variable,
at least on time scales of months, and possibly related to the luminosity of
1H~0707--495 in such a way that the features are present during higher-luminosity
states.

\subsubsection{Slim disc considerations
}
If the low-energy continuum is due to disc emission, then the measured temperature 
is too high for a standard accretion disc (discussed in T04).
It was shown in T04 that the intrinsic bolometric luminosity during the GT 
observation was near the Eddington limit 
for an
estimated mass of $\sim 2 \times 10^6 M_{\odot}$.  During the AO2 observation
the intrinsic luminosity was approximately a factor of two higher. 
In these luminosity regimes ($L_{bol} \approx 1-2 L_{edd}$) the slim-disc 
model, first proposed by Abramowicz et al. (1988), is applicable.

Mineshige et al. (2000)
adopted the slim-disc model to describe the disc emission and variability
in a sample of NLS1, including 1H~0707--495.  
A key element of the slim-disc
approximation is that the
temperature-luminosity relation is steeper when $L_{bol} \approx L_{edd}$,
than it is in the standard disc case (e.g. Mineshige et al. 2000; Watarai et al. 2001).
This can explain the observed increase in temperature against a moderate
increase in luminosity. 

A significant increase in the accretion rate
resulting in higher luminosities
could potentially explain the long-term variability seen in a warm medium
supposedly located light-months away.

\subsection{Short-term spectral variability
}
The rms spectrum during the GT observation showed no significant
spectral variability.  The explanation for this is rather straightforward
in terms of partial covering (e.g. T04). 
The AO2 spectral variability is much more complex; however, it can still be
described in the context of partial covering if we adopt a double
partial covering (Sect. 3.2; see also T04).  
A similar picture can probably be drawn for the spectral variability in
IRAS~13224--3809 (Boller et al. 2003; Gallo et al. 2004).
Admittedly, a second absorber does introduce additional parameters into the 
fit, but considering the possible complexity of partial covering in reality,
a double coverer is probably a better description of the physical processes 
involved than a single absorber is.

\subsection{Comments on previous X-ray observations of 1H~0707--495}
As indicated by its name, 1H~0707--495 was identified as a {\em HEAO1}
source.
Remillard et al. (1986) reported a 2--10~keV flux of $2.4 \times 10^{-11}$
erg cm$^{-2}$ s$^{-1}$, a factor of $\sim 20$ higher than the AO2
observation, and more than 50 times higher than the GT observation.
Clearly, source confusion was a consideration with this mission, and
it is possible that not all of the {\em HEAO1} flux is attributed 
to 1H~0707--495.  However, Remillard et al. had good reasons against a 
false detection, and no new information has surfaced to strongly
question the identification (R. Remillard 2004, priv. comm.).

About three months after the first {\em XMM-Newton} observation of
1H~0707--495, it was observed with {\em Chandra}.  A preliminary
analysis by L02 showed that the flux of 1H~0707--495
had dramatically increased by nearly a factor of 10.
Interestingly, L02 found that a $\sim 7$~keV edge was not statistically
required by the data.  The 90\% upper-limit on the edge depth was
$\tau < 0.8$.

In principle (and barring detailed analyses), these earlier observations
could be understood in the context of partial covering.  
The {\em HEAO1} flux is high (assuming it was all attributed to
1H~0707--495), but it is comparable to the intrinsic value
estimated by T04; thus it could be achieved with a near ``minimum''
in the covering fraction of the absorber. 
As with the AO2 observation, the $Chandra$ observation showing a  
diminished (possibly absent) edge and higher flux could also
arise from a reduction in the covering fraction.

\section{Conclusions} 
Two {\em XMM-Newton} observations of 1H~0707--495
separated by more than two years are examined and compared.  Our findings are as follow:
\begin{itemize}
\item[(1)]
The edge energy and depth are significantly different, while the intrinsic 
edge width remains consistently narrow ($<$ 300~eV).  Between the two 
observations, the edge energy has shifted from $\sim$ 7.0~keV to $\sim$
7.5~keV, and the edge depth has diminished by $>$ 50\%.
\item[(2)]
Emission and absorption lines are detected in the low-energy spectrum indicating
the presence of an extended warm medium.
\item[(3)]
The temperature of the MCD component used to model the soft-excess is much
higher during the AO2 observation ($kT_{AO2} \approx$ 155~eV) compared to the
GT observation ($kT_{GT} \approx$ 103~eV).
\item[(4)]
The flux variability during AO2 is typical of what has been seen in 1H~0707--495
previously.  There was also strong spectral variability during AO2
which was not observed during the GT observation.
\end{itemize}

The X-ray spectrum of 1H~0707--495 appears remarkably different over the
span of two years, showing apparent changes in the spectral slope, as well as 
changes in the edge characteristics.  We have demonstrated that
the primary X-ray source (the power-law emitter) is not required to undergo
any physical changes during this time; rather the differences can be explained simply 
by changes in a 
line-of-sight absorber associated with the partial-covering phenomenon.
The changes in the disc temperature, line-like features at low energies,
energy-dependent time variability, and shift in the edge energy may be related to
an increase of the intrinsic luminosity.  The energy-dependent time variability
can be explained assuming two separate absorbers (there could be a range of 
absorbers).  The shift of the edge energy is a challenge since it requires that the
absorber is extremely clumpy.

Further observations of 1H~0707--495 in various flux states
 should be able to 
distinguish between the partial covering and reflection models.

\section*{Acknowledgements}
Based on observations obtained with XMM-Newton, an ESA science mission with
instruments and contributions directly funded by ESA Member States and
the USA (NASA).  
We would like to thank the referee, Andrzej Zdziarski, for a critical
reading and many helpful comments.
WNB acknowledges support from NASA grant NAG5--12804.

\end{document}